# From Data Complexity to User Simplicity:
# A Framework for Linked Open Data Reconciliation and Serendipitous Discovery

Marco Grasso[1], Giulia Renda[2], Marilena Daquino[3]

[1] University of Bologna, Italy - marco.grasso7@unibo.it
[2] University of Bologna, Italy - giulia.renda3@unibo.it
[3] University of Bologna, Italy - marilena.daquino2@unibo.it

## ABSTRACT

This article introduces a novel software solution to create a Web portal to align Linked Open Data sources and provide user-friendly interfaces for serendipitous discovery. We present the Polifonia Web portal as a motivating scenario and case study to address research problems such as data reconciliation and serving generous interfaces in the music heritage domain.

## 1. INTRODUCTION

Linked Open Data (LOD) offers immense potential for data integration and knowledge discovery in the Cultural Heritage domain [2]. However, a significant challenge remains the alignment of several LOD sources while simultaneously providing user-friendly interfaces (UI) to diverse stakeholders [20].

In this article we present a reusable software solution and user interfaces we developed to address such problems, and we present the Polifonia Web portal as a case study. The Polifonia Web portal[1] addresses such challenges by providing models, methods, and interfaces tailored to (1) bridge connections within the rich and diverse domain of music heritage, and (2) foster engagement of a broad audience interested in serendipitous discovery, rather than experts' tasks only. It is designed to access a registry of music resources available online, using LOD as lingua franca to share and retrieve data. The portal offers tools to perform data aggregation and alignment, both at instance level (i.e., retrieving equivalent entities across datasets) and at ontology level (i.e. offering customizable query strategies to access sources adopting diverse ontologies). Secondly, data populate user-friendly interfaces designed to address informative needs of domain experts as well as to stimulate curiosity in the general audience, which may not have a specific task in mind.

After a concise review of the state of the art of User Interfaces for LOD-based reconciliation and exploration, we outline our methodological approach, highlight the outcomes of the research, describe the components of the proposed framework, and present the case study to validate our approach.

## 2. STATE OF THE ART

The state of the art in the field of digital data management, discovery, and interface design, particularly in the context of LOD and digital heritage, reveals several gaps and emerging challenges [10]. Existing solutions (such as *Omeka S*, *Semantic MediaWiki*, *Sinopia* and *ResearchSpace*) often fall short in key areas such as user-friendly interface building, provenance management, integration with existing data management workflows (including versioning, backup, and long-term preservation), and the reusability and sustainability of produced assets and software. These limitations are critical, as they impact the quality and sustainability of data management practices [9, 17, 19].

When addressing interfaces for displaying digital cultural heritage, research tends to split into two areas, namely: developing exploratory interfaces for the public focused on enjoyment and serendipity [7], and more complex, hypothesis-driven LOD analytical tools for experts [13]. While the concept of generous interfaces [18] aids in designing public-facing tools, effective mechanisms for LOD exploration and storytelling are notably absent. In recent works [5], 77 Linked Data visualization tools have been surveyed and evaluated. Results show that such tools are not sufficiently equipped for providing full support to users who want to explore and gain knowledge from Linked Data. Some issues regard scalability, the provision of dataset statistics beyond generic counters, and the possibility to combine different visualizations into a

---

user-defined narrative. To the best of our knowledge, no out-of-the-box solutions exist to facilitate the alignment of LOD sources and, at the same time, populate user-friendly interfaces.

## 3. APPROACH

We developed a software solution that addresses the above gaps. It facilitates the alignment of LOD sources, offering customizable data reconciliation and data indexing options, as well as the population of user-friendly interfaces including text search features and data mashup techniques to generate descriptive web pages. Our approach to designing such a web solution merges eXtreme Design [11] and Design Thinking [6, 12, 15] methodologies. Initially, the ontology design team develops personas and competency questions with input from Polifonia experts and stakeholders. Competency questions are analysed, both qualitatively and quantitatively [3, 4] to understand data and user journeys, classifying them according to their complexity and the (lack of) task at hand. A competitive analysis of existing LOD-native storytelling solutions, and web applications in cultural heritage was performed, complemented by focus groups with experts and user studies with lay users, which informed our design process, ensuring our solutions are both innovative and user-centric [8, 16]. We summarised insights to outline interaction patterns, selected deployment solutions, and conducted user testing sessions for usability validation and co-design purposes. This streamlined approach [14] ensures our design is not only advanced but also resonates well with both experts and the general public. By incorporating principles of co-design throughout this process, we ensure that our solutions are developed not just for users but with them, fostering a sense of ownership and relevance in the final product [1].

## 4. RESULTS: THE POLIFONIA WEB PORTAL FRAMEWORK

The Polifonia Web portal serves as the primary gateway to the vast array of musical heritage digital assets curated by project partners. The portal's key features include overall views of categories relevant to the domain, featured highlights to exemplify a user journey, text search functions, and the provision of on-demand insights on entities that are part of the knowledge graph. Fig.1 illustrates the main components and features that characterize the Web portal.

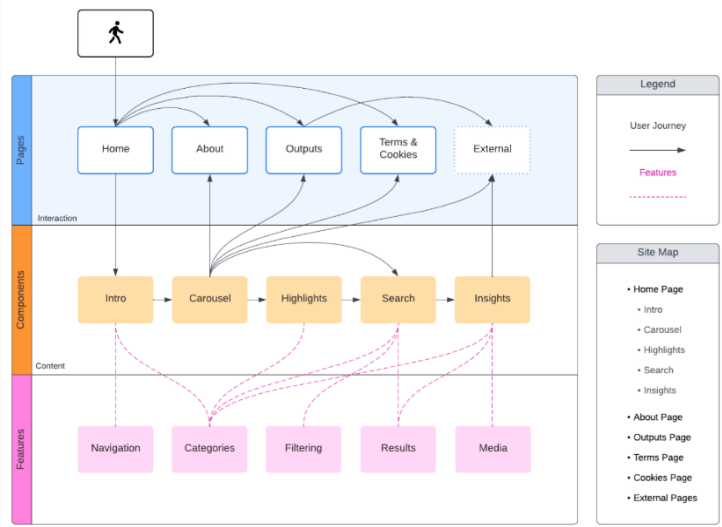

**Figure 1. Pages, components, and features of the Web portal**

The Web portal is composed of a well-orchestrated array of web pages along with a suite of interactive UI components and actionable features. The **Home page** guides users from a broad understanding of the functionalities available on the Web portal to more specific ones. The **Intro** component shows a quick take home message, and a **Carousel** shows key pages and resources connected to the Web portal (Fig. 2a).

As the user scrolls, the **Highlights** section offers a list of example dataweb available on the platform (Fig. 2b). Like a music box playing tunes, this section shows featured content and navigation options, creating a welcoming introduction to the diverse access points to the datasets indexed in the Web portal. Each highlight represents a category (in the Polifonia Web portal we adopted 5 categories, namely: genres, artists, places, music and instruments) which corresponds later in the same homepage to a section with a dedicated text search functionality. Each highlight is accompanied by a sound (which users can toggle on or off) and a colored dot. The color is encoded using the Polifonia palette and it is consistently used in subsequent sections to represent a certain category, e.g. yellow for 'places'. Highlights are displayed in a five-column layout, mirroring a music box, with each column symbolizing one of the five categories mentioned earlier. Notably, highlighted entities are linked by an associative relationship. Clicking on a highlight takes users directly to the corresponding search section.

**Search** sections allow users to explore data connected to an entity, performing a lookup search (Fig. 2c). The text search returns a list of autocompletion suggestions retrieved from an index. Each **index** includes the list of entities of the same type from different datasets, integrating cross-dataset alignments to minimize duplicates in the results. Results of the search represent relations between the search entity and other entities included in the ingested datasets. Relations between the topic and other entities are listed below the search bar, and users can filter by type of relation, category of connected entities, and source of information.

Lastly, the **Insights card** allows users to expand search results and delve into details of linked entities (Fig. 2d). A card mainly includes texts, links to external resources, multimedia, and associations between the entity at hand and other entities belonging to any of the ingested or linked datasets (e.g. Wikidata, DBpedia, Discogs).

Throughout the user journey, the intentional arrangement of components on the page ensures a seamless transition from a general overview to more specific areas of interest and searches, reflecting the original idea of creating generous interfaces. The user journey is meticulously designed to not only cater to diverse needs but also foster serendipitous discoveries by guiding users step by step through our platform's offerings and revealing information on demand. A few use cases have been designed to exemplify the potential of such an approach in real-world scenarios, namely: serendipitous discovery of multimedia connected to an artist (targeted to lay users), interconnected archives (targeted to CH curators), and music tourism guided by the influence of a music genre over nearby places (again targeted to lay users).

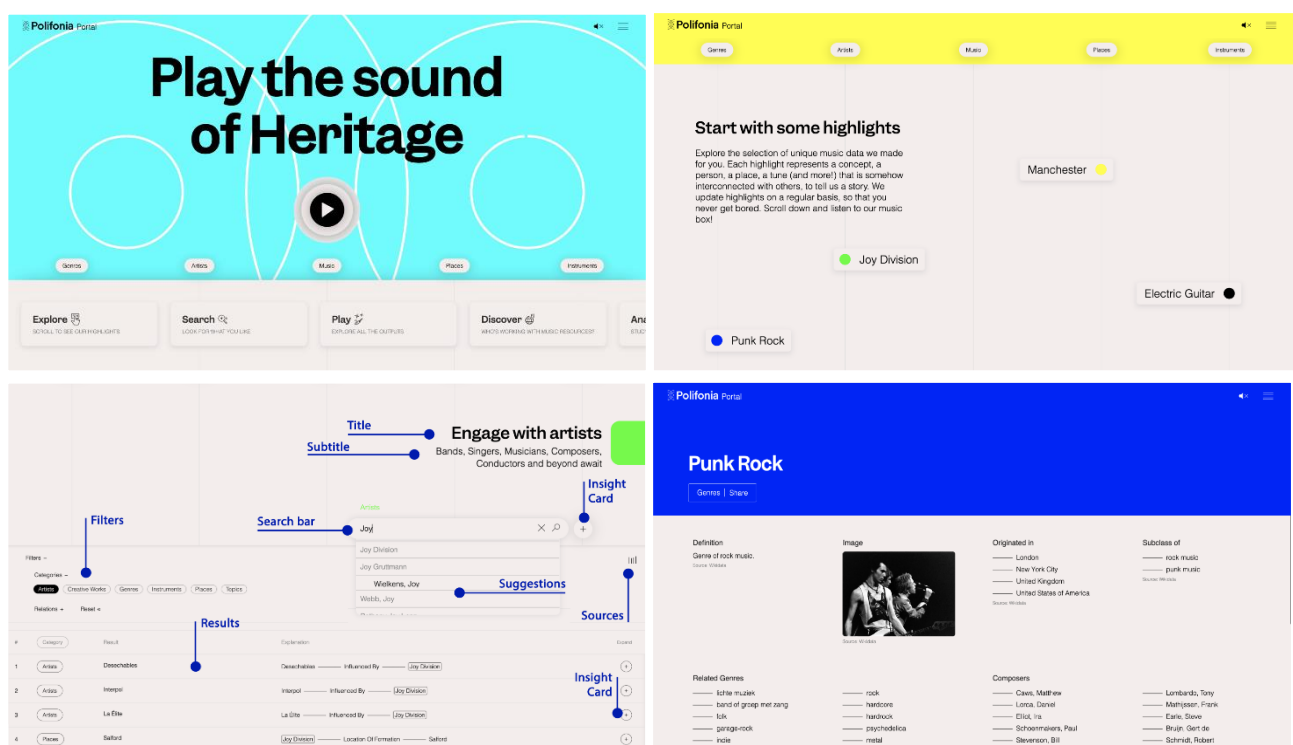

**Figure 2. The homepage of the Polifonia Web portal: Intro, Highlights, Search, Insights Card**

The Web portal's architecture is designed on top of a *Flask*[2] backend and a *React*-native[3] frontend, which communicate through **RESTful APIs**. The portal uses *Sonic*[4] for efficient data indexing, ensuring a responsive user experience (Fig. 3). The source code of the application is open source, and it is available on GitHub[5].

---

The portal is (optionally) complemented by a *Blazegraph*[6] triplestore, which plays a crucial role in storing and managing post-processed data. The data ingestion process is aimed at supporting the alignment and it begins with the extraction of data from indexed sources, as detailed in customizable **configuration files**. Such files include the SPARQL queries required to retrieve data from each dataset for each category (which are also customizable). For instance, Polifonia configuration files include SPARQL queries to retrieve all <artists>, <places>, <instruments>, <music works>, <genres> (i.e., the categories) from data sources like Wikidata, Dbpedia, etc. Once extracted, data undergoes the **reconciliation** process. This process aims to harmonize and integrate information about entities from diverse sources, contributing to solving the challenging issue of entity duplication across datasets. For example, if two different sources provide information about the same musical entity, the portal's reconciliation system will perform *sameAs* link detection, perform rule-based inference (e.g. looking for transitive links) and merge data in case of a positive match, therefore presenting the user with a unified URI and data view. The reconciliation process unfolds in several stages, starting with the creation of named graphs for each extracted URI, followed by searches for equivalence statements across Polifonia and third-party datasets like *Wikidata* and *DBpedia*. Relations are subsequently expanded leveraging direct and transitive properties and a careful lookup to identify entities appearing in multiple graphs is performed. The outcome is a **linkset** where newly minted Polifonia URIs are linked to URIs belonging to indexed data sources, which are then used to populate the indexes behind text searches.

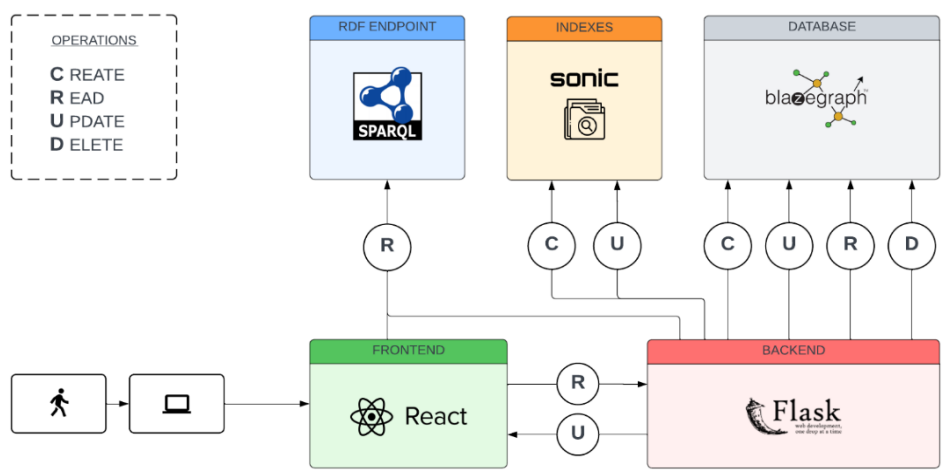

**Figure 3. Overview of the technology stack of the Web portal**

Customization and reusability are central to the Polifonia Web portal, ensuring it can easily adapt to various datasets, ontologies, and evolving UI/UX needs. This flexibility is achieved through five configuration files, enabling tailored data source access and UI component adjustments. Key customizable features include the ability to (1) add new datasets with comprehensive metadata; (2) define homepage highlights, (3) categories, (4) indexes for populating search sections, and (5) insights templates. Additionally, users can tailor the introductory section with the carousel configuration, specifying titles, descriptions, and interactive elements for each box. The most dynamic feature is the insights cards configuration, which allows for the integration of various content blocks such as text, multimedia, semantic relations, and web links. These blocks are adjustable in size, title, and description, with SPARQL queries fetching relevant content connections, offering a rich, user-centric interface.

The setup allows for the **alignment** of various datasets, making the portal not only versatile in accommodating data external to Polifonia but also **scalable** to efficiently handle new datasets and **adapt** to evolving UI requirements.

## 5. CONCLUSIONS

We have presented a prototypical web framework for data alignment and population of web pages for serendipitous discovery. Its potential impact and efficacy have been validated through multiple user studies and tests involving participants with diverse backgrounds. In total we performed four sessions divided into two phases. The initial phase was

---

performed prior to development and consisted of one session with lay users[7,8] and one with experts[9]. The second series of user tests was carried out again with general users[10,11] to better frame co-design aspects, and with experts[12,13] to test the final prototype. Feedback from these studies guided iterative improvements, ensuring that the portal not only met the technical requirements for effective LOD management but also resonated with users in terms of ease of use, navigation, and overall experience. Moreover, the tests conducted went beyond mere usability validation and delved into the practical utility of the portal in real-world scenarios. Future works will focus on assessing unresolved data quality issues in collaboration with new stakeholders and evaluating the prototype on new pilot datasets.

## 6. ACKNOWLEDGMENTS

This work is supported by a project that has received funding from the European Union's Horizon 2020 research and innovation programme under grant agreement No 101004746 (Polifonia: a digital harmoniser for musical heritage knowledge, H2020-SC6-TRANSFORMATIONS).